\documentclass[conference]{IEEEtran}
\IEEEoverridecommandlockouts
\usepackage{cite}
\usepackage{amsmath,amssymb,amsfonts}
\usepackage{algorithmic}
\usepackage{graphicx}
\usepackage{textcomp}
\usepackage{xcolor}
\usepackage{orcidlink}
\usepackage{lipsum}
\usepackage{stfloats} %
\usepackage{booktabs, tabularx, pifont, makecell, xcolor, colortbl}
\def\BibTeX{{\rm B\kern-.05em{\sc i\kern-.025em b}\kern-.08em
    T\kern-.1667em\lower.7ex\hbox{E}\kern-.125emX}}

\newcommand{\footurl}[1]{\footnote{\url{#1}}}
\newcommand{\Builder}{\textsc{DesCartes Builder}}

\begin{document}

\title{\Builder: A Tool to Develop Machine-Learning Based Digital Twins
\thanks{This research is part of the program DesCartes and is supported by the National Research Foundation, Prime Minister's Office, Singapore under its Campus for Research Excellence and Technological Enterprise (CREATE) program.}
}

\author{\IEEEauthorblockN{1\textsuperscript{st} Eduardo de Conto \orcidlink{0009-0003-9217-0890}}
\IEEEauthorblockA{
    \textit{Nanyang Technological University} \\
    \textit{CNRS@CREATE} \\
    Singapore, Singapore
}
\and
\IEEEauthorblockN{2\textsuperscript{nd} Blaise Genest \orcidlink{0000-0002-5758-1876}}
\IEEEauthorblockA{\textit{IPAL Lab}\\
\textit{CNRS, CNRS@CREATE}\\
France and Singapore
}
\and
\IEEEauthorblockN{3\textsuperscript{rd} Arvind Easwaran \orcidlink{0000-0002-9628-3847}}
\IEEEauthorblockA{
    \textit{Nanyang Technological University}\\
    Singapore, Singapore
}
\and 
\IEEEauthorblockN{4\textsuperscript{th} Nicholas Ng}
\IEEEauthorblockA{\textit{CNRS@CREATE}\\
Singapore, Singapore
}
\and
\IEEEauthorblockN{5\textsuperscript{th} Shweta Menon}
\IEEEauthorblockA{
    \textit{CNRS@CREATE}\\
    Singapore, Singapore
}
}

\maketitle

\begin{abstract}
Digital twins (DTs) are increasingly utilized to monitor, manage, and optimize complex systems across various domains, including civil engineering. A core requirement for an effective DT is to act as a fast, accurate, and maintainable surrogate of its physical counterpart, the physical twin (PT). 
To this end, machine learning (ML) is frequently employed to
(i) construct real-time DT prototypes using efficient reduced-order models (ROMs) derived from high-fidelity simulations of the PT's nominal behavior, and
(ii) specialize these prototypes into DT instances by leveraging historical sensor data from the target PT. 
Despite the broad applicability of ML, its use in DT engineering remains largely ad hoc. Indeed, while conventional ML pipelines often train a single model for a specific task, DTs typically require multiple, task- and domain-dependent models. Thus, a more structured approach is required to design DTs.

In this paper, we introduce \textsc{DesCartes Builder}, an open-source tool to enable the systematic engineering of ML-based pipelines for real-time DT prototypes and DT instances. The tool leverages an open and flexible visual data flow paradigm to facilitate the specification, composition, and reuse of ML models. It also integrates a library of parameterizable core operations and ML algorithms tailored for DT design. We demonstrate the effectiveness and usability of \textsc{DesCartes Builder} through a civil engineering use case involving the design of a real-time DT prototype to predict the plastic strain of a structure.

\end{abstract}


\bstctlcite{IEEEexample:BSTcontrol}

\section{Introduction}

\emph{Digital twins (DTs)} are software entities that accurately mirror and co-evolve with their corresponding physical counterpart, the physical twin (PT). DTs are increasingly deployed in safety-critical domains, such as civil engineering, to enable, e.g., preventive maintenance and continuous performance optimization. 
Despite their significance, the systematic engineering of DTs remains largely ad hoc. %
While traditionally DTs were designed using high-fidelity, physics-based methods, these methods quickly become computationally expensive and numerically unstable as the DT complexity increases. %
Consequently, data-driven predictors based on machine learning (ML) are gaining traction, enabling fast and scalable predictions~\cite{fouquetGreyCatFrameworkDevelop2024, chinestaVirtualDigitalHybrid2020}.
ML is typically used for two key tasks: 
\begin{itemize}
    \item \emph{Reduced-Order Modeling for Real-Time DT Prototypes}. Given a high-fidelity simulation of the PT nominal behavior (a DT prototype), ML, specifically \emph{unsupervised learning} techniques such as principal component analysis (PCA), is used to obtain efficient \emph{reduced-order models (ROMs)} of the PT. This results in a \emph{real-time DT prototype}, which enables significantly faster simulations without sacrificing accuracy~\cite{hartmannModelOrderReduction2018}.

    \item \emph{Data Assimilation for DT Instances}. The real-time DT prototype is \emph{specialized} for a particular PT instance, resulting in a \emph{DT instance}, which accounts for, e.g., manufacturing defects. This process involves \emph{data assimilation}, often based on supervised learning (e.g., neural networks trained through gradient descent) using historical sensor data as the supervised input~\cite{donatoSelfupdatingDigitalTwin2024}. 
\end{itemize}

However, while the main ROM and instance-specialization building blocks and methodologies are generally applicable, the pipelines connecting these components are highly use-case dependent. 
Thus, the application of conventional \emph{ML workflow tools} (e.g., Kedro~\cite{Kedro}) is challenging, since they often train a single, task-specific ML model, which is not explicitly represented in the pipeline. %

\begin{figure*}[t!]
    \centering
    \includegraphics[width=.93\textwidth]{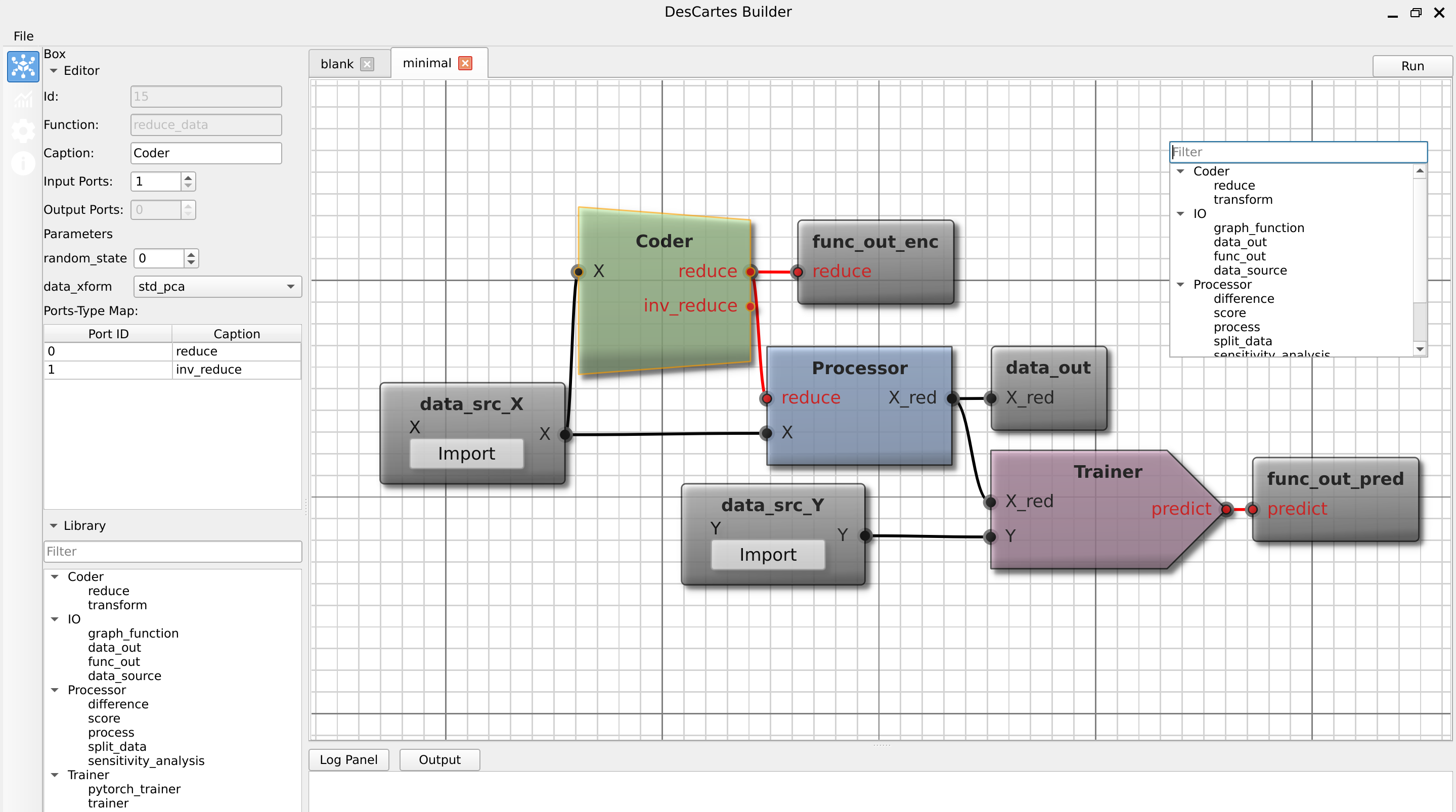}
    \caption{{\Builder} user interface showing the canvas, the library selector, and the parameter editor. A minimal FDF pipeline is displayed:
    An unsupervised ML Coder computes a reduced space for the input data \texttt{X}, returning its associated {\color{red}\texttt{reduce}} function;
    a Processor encodes the \texttt{X} as \texttt{X\_red} in this reduced space by applying {\color{red}\texttt{reduce}}; 
    a supervised ML Trainer learns a {\color{red}\texttt{predict}} function that takes \texttt{X\_red} and outputs an estimate of the DT simulation \texttt{Y}, with (\texttt{X\_red}, \texttt{Y}) serving as supervised inputs. 
    The functions {\color{red}\texttt{predict}} and {\color{red}\texttt{reduce}} are exported to constitute the real-time DT prototype.}
    \label{fig:builder-startup}
\end{figure*}

To our knowledge, no existing tool supports the design of real-time DT prototypes and DT instances using a \emph{flexible workflow} in which multiple models can be explicitly manipulated (created, used, reused) in the pipeline. This makes DT development challenging, in particular for \emph{domain experts}, users that typically lack programming expertise and are also not interested in the intricacies of the underlying models.

In an attempt to fill this gap, we introduce {\Builder}. The tool leverages model-driven engineering (MDE) principles, which can render DT development more broadly accessible across its various phases~\cite{lehnerModeldrivenEngineeringDigital2025}. %
Our key \emph{contributions} are as follows:
\begin{itemize}
    \item We publicly release {\Builder}, a tool aiming to democratize the synthesis and validation of DTs~\cite{decontoDesCartesBuilderDigital2025}. The tool enables users (specifically, domain experts) to:
    \begin{itemize}
        \item model the DT synthesis pipeline graphically, including explicit modeling of data flows, function flows, and their interconnections,
        \item execute the DT synthesis pipeline, obtaining data artifacts and ML models constituting the DT.
    \end{itemize}
    \item We describe the tool's architecture and implementation. 
\end{itemize}

\section{\Builder~Overview}

{\Builder} implements the Function+Data Flow (FDF) domain-specific language for ML-based DT design~\cite{decontoFunction+DataFlowFramework2024}.
FDF employs a high-order visual data flow paradigm in which functions (representing ML models) are first-class citizens: they can be both inputs and outputs of FDF's processing units (\emph{boxes}). There are three core FDF boxes:
\begin{itemize}
    \item \emph{Processor}, represented by a light blue rectangle, for reusing functions learned by other boxes, as well as predefined functions.
    \item \emph{Coder}, represented by a pale green trapezoid, for learning an encoding/decoding function with unsupervised ML,
    \item \emph{Trainer}, represented by a pale violet pentagon, for learning a function with supervised ML.
\end{itemize}

In the following, we describe the tool's architecture, which is structured into two primary components: a graphical front end and an execution back end. This ensures a clear separation between the user interface (UI) and the core computational logic. The interaction between these components is mediated by an \emph{abstract execution engine}, providing a decoupled communication layer. {\Builder} is \emph{open-source}, supporting extensibility and community-driven development~\cite{decontoDesCartesBuilderDigital2025}.

\begin{figure*}[t!!]
    \centering
    \includegraphics[width=0.93\linewidth]{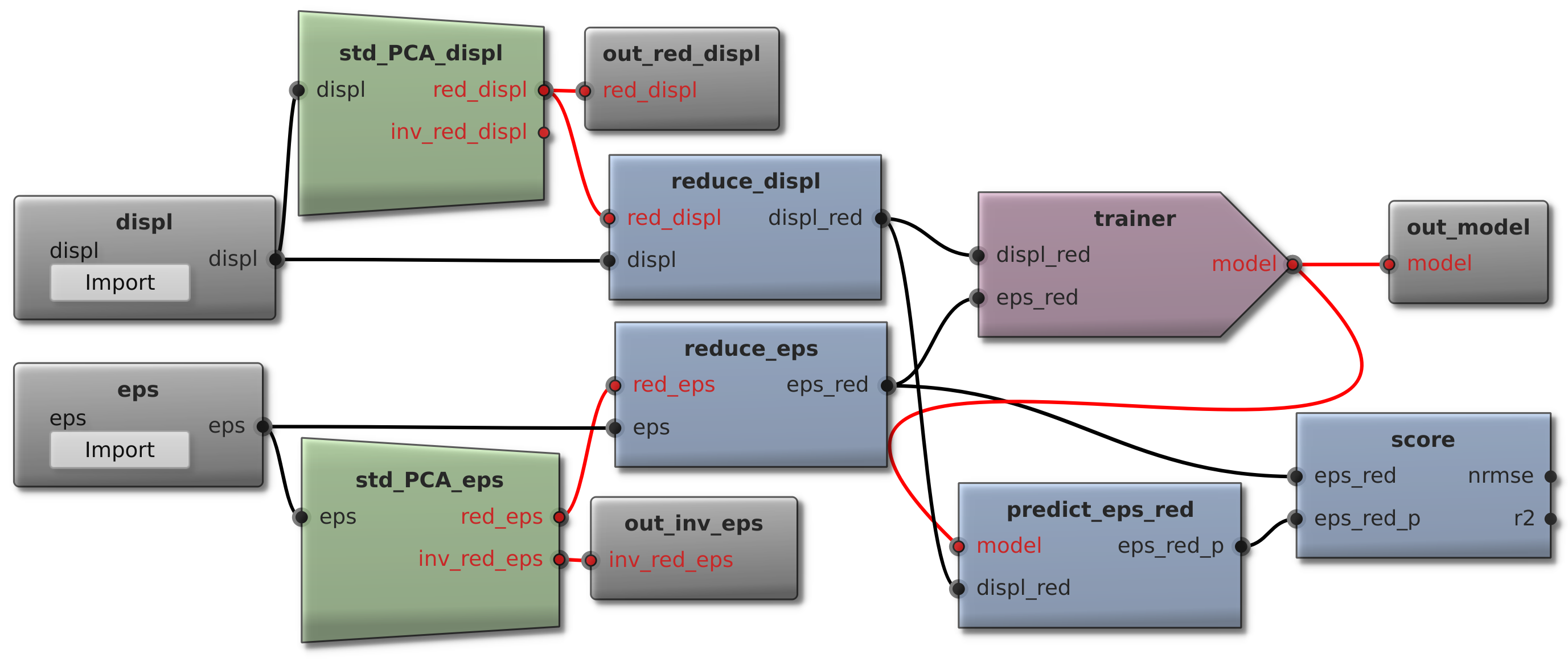}
    \caption{Implementation of the material strain ROM. It uses two Coder boxes to learn the dimensionality reduction basis, two Processors to obtain the reduced displacement (\texttt{displ\_red}) and reduced strain (\texttt{eps\_red}), and a Trainer to learn a surrogate model from \texttt{displ\_red} to \texttt{eps\_red}. Evaluation is performed using a \texttt{score} Processor box.}%
    \label{fig:pipe-deformation-train}
\end{figure*}

\subsection{Front End User Interface}

{\Builder} front end offers a unified graphical UI for specifying, executing, and validating ML-based DT pipelines. Implemented in C++ using the Qt framework and the QtNodes library~\cite{pinaevQtNodesNodeEditor2024}, it offers a responsive, cross-platform user experience. The UI, depicted in Fig.~\ref{fig:builder-startup}, comprises the following core components:
\begin{enumerate}
    \item a central \emph{canvas} for specifying DT pipelines using FDF, allowing users to create boxes and to interconnect their ports through drag-and-drop. Function and data connections are depicted in \textbf{\color{red}red} and \textbf{black}, respectively, for clear differentiation. 
    \item a \emph{library selector} (canvas pop-up and left pane) and a \emph{parameter editor} (left pane) for inserting FDF boxes and modifying their parameters (e.g., adjust the data encoding learned by the Coder), respectively.
    \item A \texttt{Run} button (top-right) for triggering the \emph{pipeline execution}.
    \item A \verb|Log Panel| and \verb|Output| panel (bottom) for displaying execution traces.
    \item A \emph{chart viewer} (\includegraphics[height=0.8\baselineskip]{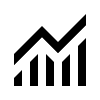}) for visualizing charts and validation metrics by selecting appropriate boxes (e.g., \verb|score| Processor).
\end{enumerate}

The \emph{pipeline execution} proceeds by first generating code for the back end based on the FDF boxes and their associated parameters. It then launches the back end and waits for the DT model generation to complete, training and validating ML models as needed. Upon completion, the results are collected and presented in both the \verb|Output| panel and the \emph{chart viewer}. This process is coordinated by the \emph{abstract execution engine}.

\subsection{Back End Execution Engine}

The back end, implemented in Python as Kedro~\cite{Kedro} plugin, has two main responsibilities:
\begin{enumerate}
    \item \emph{Pipeline Orchestration}. Executes the pipeline according to the FDF semantics. This involves loading the input data, determining the execution order and dependencies within the FDF graph, and saving the results as required. 
    \item \emph{Library Definition}. Provides a library of predefined functions and learning algorithms. Each front-end box is implemented as a function in this library. 
\end{enumerate}

\section{Real-Time DT Prototype for Material Strain Prediction}
\label{sec:dtp-material-strain}

We now demonstrate the application of {\Builder} to design a real-time DT prototype for predicting the material strain of a civil infrastructure. This DT can be used for structural health monitoring, enabling predictive maintenance. 
While the plastic strain cannot currently be measured non-destructively, the material deformation can be estimated using, e.g., high-resolution photos of the material~\cite{johnsonComplexitiesCapturingLarge2023}.

To design such a DT, we leverage high-fidelity simulations of the system obtained using a finite element model (FEM) of the structure and shared with us by a collaborator. These simulations accurately estimate the plastic strain (\verb|eps|) and deformation (\verb|displ|) for different parameters of the material (impact strength, thickness of the material, etc.). However, since each simulation requires around one hour of runtime, they cannot be used directly to build the real-time DT prototype. To address this,  we employ a design of experiments (DoE) strategy to systematically sample parameter combinations covering all meaningful behaviors of the material. 

Using this DoE data, we obtain a ROM of the structure in {\Builder}, as shown in Fig.~\ref{fig:pipe-deformation-train}. The pipeline has three main steps: 
\begin{itemize}
    \item \emph{Dimensionality Reduction}. We learn a transformation reducing the dimensionality of the deformation and strain meshes from $>1000$ dimensions to $\approx10$ dimensions. This is achieved with two Coder boxes (\verb|std_PCA_displ| and \verb|std_PCA_eps|), which use standardization followed by PCA. To obtain the reduced data, two Processor boxes (\verb|reduce_displ| and \verb|reduce_eps|) encode the full-dimensional data with the encodings \verb|red_displ| and \verb|red_eps| learned by the Coder boxes, resulting in \verb|displ_red| and \verb|eps_red|, the reduced displacement and plastic strain. 
    \item \emph{Surrogate Learning}. We learn a surrogate to predict \verb|eps_red| from \verb|displ_red| using the dataset obtained in the previous step. This is achieved using a Trainer box, parameterized to learn a neural network by gradient descent, and outputting the resulting \verb|model|.
    \item \emph{Surrogate Evaluation}. We evaluate the learned surrogate \verb|model| using a \verb|score| Processor box, which compares displacement predictions (\verb|eps_red_p|) against ground truth values (\verb|eps_red|). Results are visualized in the \emph{chart viewer} (Fig.~\ref{fig:pipe-deformation-score}). For illustration, evaluation is performed on the training set here; however, more robust evaluation procedures may be necessary in practice.
\end{itemize}

{\Builder} facilitates the modeling of this pipeline in three ways. First, it allows describing visually the data flow necessary to learn a function. Second, it enables the easy export of functions using the gray function output boxes (\verb|out_red_displ|, \verb|out_inv_eps|, and \verb|out_model|). Third, it supports FDF's implicit typing, which can identify misconnections at specification time (Fig.~\ref{fig:implicit-typing}).

\section{Related Works}

\newcommand{\cmark}{\textcolor{green!80!black}{\ding{51}}}
\newcommand{\xmark}{\textcolor{red}{\ding{55}}}

\begin{table}[h]
    \centering
    \caption{Comparison of features for DT engineering between {\Builder} and related work. \cmark = fulfilled, \xmark = not fulfilled.}
    \label{tab:tool-comparison}
    \begin{tabular}{l*{6}{l}}
        \toprule
        Features                             & \rotatebox{90}{Graphical Modeling UI}
                                             & \rotatebox{90}{Customizable Workflow}
                                             & \rotatebox{90}{ML Training}
                                             & \rotatebox{90}{Built-in Library}
                                             & \rotatebox{90}{Open-Source}
                                             & \rotatebox{90}{Domain Expert Friendly}                                                                                                                                             \\ \midrule
        Ansys Twin Builder & \cmark & \xmark & \xmark & \cmark & \xmark & \cmark \\
        Azure Digital Twins or GreyCat & \xmark & \cmark & \xmark & \cmark & \xmark & \xmark \\
        Kedro or MLFlow & \xmark & \cmark & \cmark & \xmark & \cmark & \xmark \\
        \cellcolor{gray!30}\Builder & \cellcolor{gray!30}\cmark & \cellcolor{gray!30}\cmark & \cellcolor{gray!30}\cmark & \cellcolor{gray!30}\cmark & \cellcolor{gray!30}\cmark & \cellcolor{gray!30}\cmark \\
        \bottomrule
    \end{tabular}
\end{table}

A broad range of tools can be leveraged for the development of real-time DT prototypes and DT instances, as summarized in Table~\ref{tab:tool-comparison}. 
Traditionally, ROMs for real-time DT prototypes are obtained using tools from commercial vendors of physics-based modeling tools, such as Ansys and Siemens. The workflow is implemented within proprietary platforms (e.g., Twin Builder~\cite{AnsysTwinBuilder} and Simcenter Amesim~\cite{SimcenterSystemsSimulation}), and it is often predefined, with limited user-controllable parameters. Consequently, the results depend significantly on whether the available workflows are suitable for the domain of interest. {\Builder} addresses these limitations by providing an \emph{open-source}, fully \emph{customizable workflow} for reduced-order modeling and instance-specialization using FDF.

To obtain the more flexible, data-driven models required for DT design, open-source ML workflow tools (e.g., Kedro~\cite{Kedro} and MLFlow~\cite{mlflow}) can be leveraged. These tools enable the systematic representation of complex ML pipelines, facilitating experiment tracking, model versioning, and performance monitoring. Nonetheless, these frameworks primarily focus on a single ML model, which is not explicitly represented in the pipeline. Moreover, their interfaces are primarily textual, centered around user-defined code. In contrast, {\Builder} features a \emph{graphical modeling UI} based on FDF. This enables an intuitive and adaptable pipeline description that is \emph{friendly to domain experts}, and in which FDF's dedicated function flow enables the explicit manipulation of the various models constituting the DT. 
In addition, {\Builder} includes a \emph{built-in library} with key algorithms and ML methods for DT engineering. 

Modeling DTs also involves effectively representing and interacting with the PT. 
For example, in the material strain prediction use case, it may be essential to capture that a material is part of a particular civil infrastructure and that multiple such materials need to be monitored simultaneously. Platforms such as Azure Digital Twins~\cite{AzureDigitalTwins} and GreyCat~\cite{fouquetGreyCatFrameworkDevelop2024} emphasize data definition and facilitate DT-PT integration. However, they do not have built-in support for ML training, instead supporting only online ML inference. In contrast, {\Builder} enables \emph{ML training} and it can easily export the ML models required for PT interaction.

\section{Conclusion}

In this paper, we presented {\Builder}, an open-source tool to develop ML-based DTs, specifically for model-order reduction and instance-specialization. The tool adopts an intuitive visual data flow paradigm aimed at domain experts with minimal programming expertise. We detailed its architecture, which leverages open-source components to promote extensibility and reusability. 
To illustrate its capabilities, we applied {\Builder} to a plastic strain prediction use case, demonstrating how it facilitates the specification and training of ML-based DTs within a unified environment.

\bibliographystyle{IEEEtran}  
\bibliography{refs}

\clearpage
\appendix[Demonstration Outline]
{\Builder} documentation, tool screenshots, and a Docker container for running the tool are available online~\cite{decontoDesCartesBuilderDigital2025}. The development version is hosted on GitHub: \href{https://github.com/CPS-research-group/descartes-builder}{front end}, \href{https://github.com/CPS-research-group/kedro-umbrella}{back end}.
A short video demonstration is available at \url{https://youtu.be/MTaR3ewBuBQ}.

\begin{figure*}[b]
    \centering
    \includegraphics[width=0.95\textwidth]{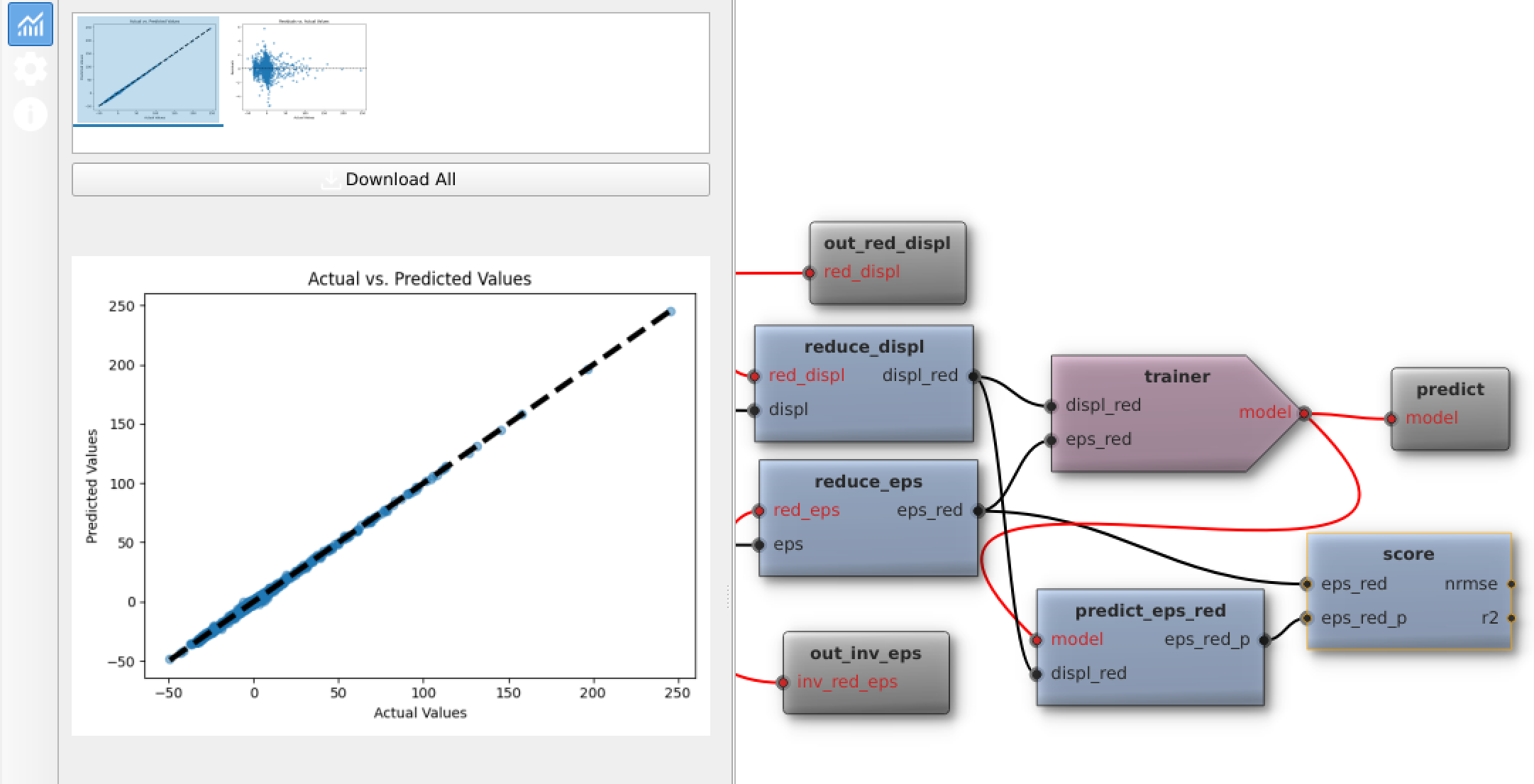}
    \caption{Validation results for the material strain prediction pipeline are displayed in the chart viewer when the \texttt{score} Processor box is selected. Among others, an actual vs predicted value plot is displayed. The plot indicates a high predictive accuracy, since all points are close to the ideal $45^\circ$ line, indicating strong agreement between the supervised ML target (\texttt{eps\_red}) and model output (\texttt{eps\_red\_p}) when \texttt{displ\_red} is used as input.}
    \label{fig:pipe-deformation-score}
\end{figure*}

\begin{figure*}[b]
    \centering
    \includegraphics[width=0.95\textwidth]{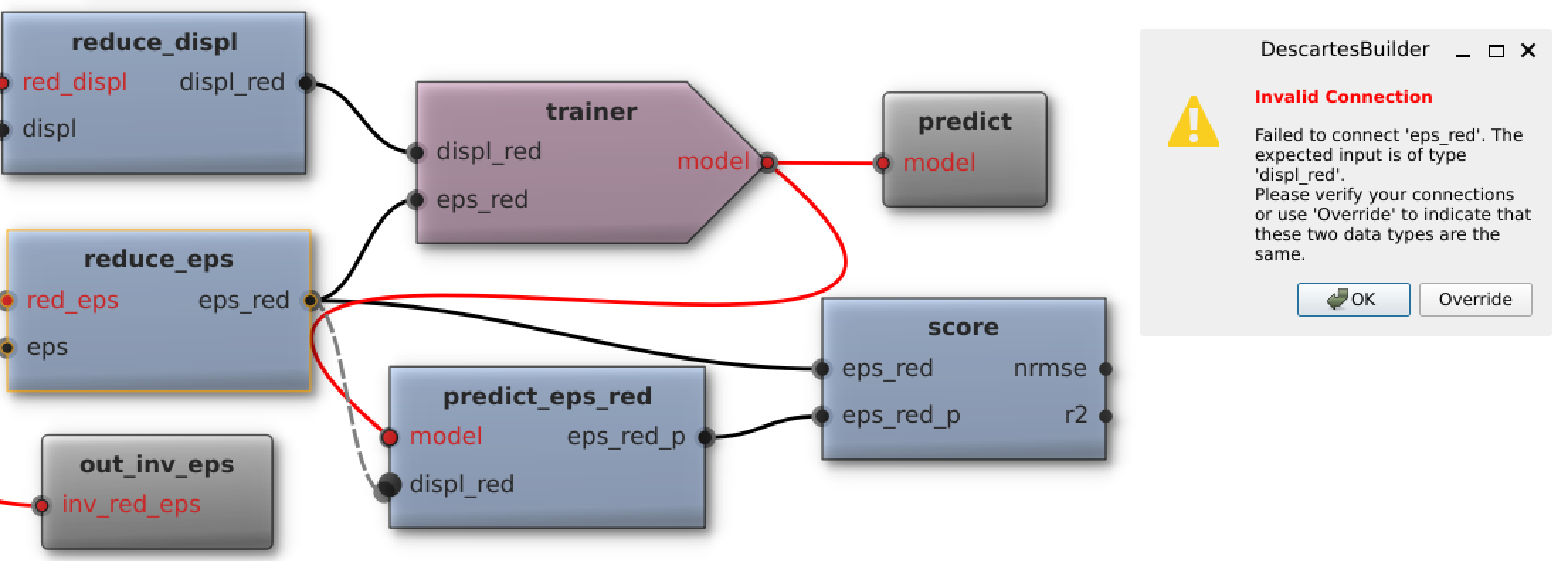}
    \caption{
    An implicit typing warning is raised at \emph{specification time} when \texttt{eps\_red} is connected where \texttt{displ\_red} is expected. This occurs because the \texttt{model} function (produced by the Trainer block) expects an input matching \texttt{displ\_red}'s implicit type, but instead receives \texttt{eps\_red} when reused on the \texttt{predict\_eps\_red} block. To address this, the user can either fix the pipeline or use the \texttt{Override} option to indicate that \texttt{displ\_red} and \texttt{eps\_red} share the same implicit type.}
    \label{fig:implicit-typing}
\end{figure*}

\end{document}